\documentclass[journal]{IEEEtran}

\ifCLASSINFOpdf
\else
   \usepackage[dvips]{graphicx}
\fi
\usepackage{url}

\usepackage{graphicx}
\usepackage{float}
\usepackage{algorithmic}
\usepackage{subfigure,multirow,bm,stfloats}
\usepackage{amsmath,amssymb,amsfonts}

\usepackage{algorithm}
\usepackage{xcolor}
\usepackage{soul}
\begin{document}

\title{Image Processing via Multilayer Graph Spectra}

\author{Songyang Zhang, Qinwen Deng, and Zhi Ding, \IEEEmembership{Fellow, IEEE}
\thanks{S. Zhang, Q. Deng, and Z. Ding are with the Department of Electrical and Computer
	Engineering, University of California at Davis, Davis, CA 95616 USA (e-mail:
	sydzhang, mrdeng, and zding@ucdavis.edu).}
}

\maketitle

\begin{abstract}
Graph signal processing (GSP) has become an important tool in image processing because of its ability to reveal underlying data structures. Many real-life multimedia datasets, however, exhibit heterogeneous structures across frames. Multilayer graphs (MLG), instead of traditional single-layer graphs, provide better representation of these
datasets such as videos and hyperspectral images. To generalize GSP to multilayer graph models and develop multilayer analysis for image processing, this work introduces a tensor-based framework of multilayer graph signal processing (M-GSP) and present useful M-GSP tools for image processing. We then present guidelines for applying M-GSP in image processing and introduce several applications, including RGB image compression, edge detection and hyperspectral image segmentation. Successful experimental results demonstrate the efficacy and promising futures of M-GSP in image processing.
\end{abstract}


\IEEEpeerreviewmaketitle

\section{Introduction}

\begin{figure}[t]
	\centering
	\includegraphics[width=2in]{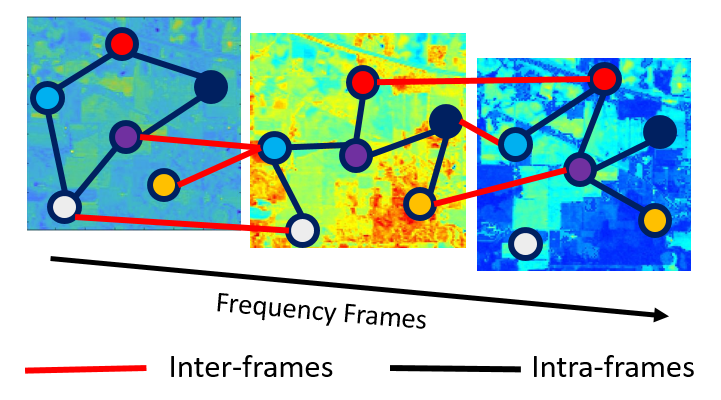}
	\vspace{-2mm}
	\caption{Example of Multilayer Graph Model for Hyperspectral Images: pixels are represented by nodes and their similarities are represented by edges.}
	\label{exm}
	\vspace{-5mm}
\end{figure}

\IEEEPARstart{S}{ignal} processing over graphs has demonstrated successes in image processing \cite{c1} owing to its ability to uncover
some hidden data features. Representing pixels (or superpixels) by nodes and their internal interactions by edges, graphs can be formed to describe the underlying image structures. 
In graph signal processing (GSP),
graph Fourier space \cite{c2} and the corresponding spectral analysis have been
applied in 
image compression \cite{c3} and denoising \cite{c4}. Nevertheless, 
single-layer graphs tend to be less efficient
in processing high-dimensional images, where each frame corresponds to a distinct
geometric structure. For example,  hyperspectral images at different frequencies may exhibit heterogeneous graph structures, and do not lend themselves to traditional GSP tools. Similar scenarios include the color frames in RGB images and temporal frames in videos. On the other hand, 
since these frames are correlated, their inter-frame correlations should not be ignored in graph representations. Such complex intra- and inter- frame interactions among pixels can be represented by multilayer graphs (or multilayer networks) instead of a single-layer graph. 
Shown as Fig. \ref{exm}, a hyperspectral image can be naturally modeled as a multilayer graph (MLG) with each pixel position
as a node and each frequency frame as one layer. Of strong interest are
questions of how to generalize single-layer GSP to multilayer graph signal processing (M-GSP) and how to extract additional benefits of M-GSP in problems such as image processing.

Graph signal processing over multilayer networks (graphs) has attracted increasing attentions recently. In \cite{c6}, a two-step graph Fourier transform (GFT) is separately implemented spatially and temporally to compress spatial-temporal skeleton data. Similarly, authors of \cite{c7}
defined a joint time-vertex Fourier transform (JFT) by applying GFT and DFT consecutively on
time-varying datasets. However, both the
two-step GFT and JFT assume that the spatial connections in all layers match the same underlying graph structure, thereby limiting the use of such spectral analysis on  heterogeneous data frames. 
In addition to matrix-based GSP analysis, a tensor-based multi-way graph signal processing framework (MWGSP), proposed in \cite{c8},
decomposes the high-dimensional tensor signal into individual orders and constructs factor graphs for each order. 
Although MWGSP allows more complex interlayer connections and is suitable for high-dimensional signals, 
it assumes the dataset to lie in a product graph from all factor graphs, therefore
still requires all intralayer connections 
to exhibit homogeneous structures from the node-wise factor graph. 

To capture the heterogeneous layer structures in MLG for generalized GSP, a framework of multilayer network signal processing (M-GSP) has been proposed by using tensor representations \cite{c9}. Defining MLG spectra via tensor decomposition, both joint and order-wise analysis for MLG can
tackle various intralayer and complex interlayer connections. However, we note that existing M-GSP tools still lack some definitions of basic MLG spectral operations useful in image processing, such as sampling and convolution.

In this work, we investigate the use of M-GSP spectral analysis \cite{c9} and
the use of M-GSP tools in image processing. The contributions can be summarized as follows:
\begin{itemize}
	\item We first introduce the MLG models and representations for images, and then define several M-GSP tools useful for image processing.
	\item We provide a tutorial on applying M-GSP in image processing and present several practical cases, including image compression, edge detection, and hyperspectral image segmentation.
\end{itemize}
We compare each proposed M-GSP image 
application to benchmarks 
from traditional GSP and/or
learning methods. Our
test results demonstrate the strength of M-GSP and the proposed MLG spectral tools in 
processing images.

\section{Models and Spectrum of Images in M-GSP}\label{ed}
In this section, we introduce the MLG model of images and
outline fundamentals of M-GSP \cite{c9}. 
We refrain from using the term “network”
because of its different meanings in various field ranging from communications to deep learning. From here onward, unless otherwise specified, we shall use the
less ambiguous term of multilayer graph  instead.

To model related image frames as multilayer graph (MLG), we can naturally define layers as frames and nodes as pixels (superpixels) in each frame. Constructing interactions based on feature similarity or physical connections,
a multilayer graph with the same number of nodes in each layer can be constructed to represent the images. Moreover, viewing $N$ (super)pixels $\{x_1,\cdots,x_N\}$ as entities, the MLG structure can be viewed as projecting the physical entities into 
natural representations of $M$ frames $\{l_1,\cdots,l_M\}$, e.g., spectrum band frames for hyperspectral images, color frames for RGB images, and temporal frames for time-varying images. Such multilayer graph structure with $M$ layers and $N$ nodes in each layer can be intuitively represented by a forth-order adjacency tensor $\mathbf{A}=(A_{\alpha i \beta j})\in \mathbb{R}^{M\times N\times M \times  N}$ with $1\leq \alpha,\beta\leq M, 1\leq i,j\leq N$.
Here, each entry $A_{\alpha i \beta j}$ of the adjacency tensor $\mathbf{A}$ indicates the  edge strength between entity $j$'s projected node in layer $\beta$ and entity $i$'s projected node in layer $\alpha$. Similar to normal graphs, Laplacian tensor $\mathbf{L}\in \mathbb{R}^{M\times N\times M \times  N}$ can be defined as the alternative representation of MLG. Interested readers may refer to \cite{c9} for more details of the adjacency and Laplacian tensor. For convenience, we use a general notation $\mathbf{F}\in \mathbb{R}^{M\times N\times M \times  N}$ to represent the MLG in this work. In addition to the representation of MLG, signals over MLG can be intuitively defined as $\mathbf{s}=({s_{\alpha i}})\in\mathbb{R}^{M\times N}$, with $s_{\alpha i}$ as the value of pixel $i$ in frame $\alpha$.

In an undirected multilayer graph, the
representing tensor (adjacency/Laplacian) $\mathbf{F}$ is partially
symmetric between orders one and three, and 
between orders two and four, respectively. Then, it can be approximated via orthogonal CANDECOMP/PARAFAC (CP) decomposition \cite{c10} as
\begin{align}\label{decompose1}
\mathbf{F}&\approx\sum_{\alpha=1}^{M}\sum_{i=1}^N \lambda_{\alpha i} \cdot \mathbf{f}_\alpha \circ\mathbf{e}_i\circ \mathbf{f}_\alpha \circ\mathbf{e}_i,
\end{align}
where $\circ$ is the tensor outer product, $\mathbf{f}_\alpha \in\mathbb{R}^M$ are orthonormal, $\mathbf{e}_i\in\mathbb{R}^N$ are orthonormal and $\tilde{\mathbf{V}}_{\alpha i}=\mathbf{f}_{\alpha}\circ \mathbf{e}_i\in\mathbb{R}^{M\times N}$. Here, $\mathbf{f}_\alpha$ and $\mathbf{e}_i$ are named as the MLG spectral bases characterizing the features of layers and entities, respectively.

Besides MLG spectral space, the singular space is defined from HOSVD \cite{c11} as an alternative subspace of MLG, i.e.,
\begin{equation}\label{decomposeS}
\mathbf{F}= \mathbf{S}\times_1 \mathbf{U}^{(1)}\times_2 \mathbf{U}^{(2)}\times_3 \mathbf{U}^{(3)}\times_4 \mathbf{U}^{(4)},
\end{equation}
where $\times_n$ denotes the $n$-mode product \cite{c12} and $\mathbf{U}^{(n)}$ is a unitary $(I_n\times I_n)$ matrix, with $I_1=I_3=M$ and $I_2=I_4=N$. For an undirected multilayer graph, the adjacency tensor is symmetric for
every 2-D combination. Thus, there are two modes of singular spectrum, i.e., $(\gamma_\alpha, \mathbf{f}_\alpha)$ for mode $1,3$, and $(\sigma_i,\mathbf{e}_i)$ for mode $2,4$. 
More specifically, $\mathbf{U}^{(1)}=\mathbf{U}^{(3)}=(\mathbf{f}_\alpha)$ and $\mathbf{U}^{(2)}=\mathbf{U}^{(4)}=(\mathbf{e}_i)$. Singular tensor analysis and spectral analysis are both efficient tools for image processing depending on specific tasks. We shall elaborate in Section \ref{app}.

Suppose that $\mathbf{E}_f=[\mathbf{f}_1,\cdots,\mathbf{f}_M]$ and $\mathbf{E}_e=[\mathbf{e}_1,\cdots,\mathbf{e}_N]$ include all the layer-wise (frame) and entity-wise (pixel) spectral (or singular) vectors. The MLG Fourier (or Singular) transform can be defined as
$\hat{\mathbf{s}}=\mathbf{E}_f^{\mathrm{T}}\mathbf{s}\mathbf{E}_e\in\mathbb{R}^{M\times N}$, and the inverse transform is given as 
$\mathbf{s}'=\mathbf{E}_f\hat{\mathbf{s}}\mathbf{E}_e^{\mathrm{T}}$. Due to page limitation, here we only introduce fundamentals of MLG spectral and singular vectors for MLG analysis. More details, such as spectral interpretation, frequency analysis, and filter design can be found in \cite{c9}.

\section{M-GSP Tools for Image Processing}
In this section, we introduce important operations for M-GSP frequency analysis 
specially useful in image processing.
\subsubsection{MLG Spectral Convolution}
Graph-based spectral convolution is useful in applications 
such as feature extraction, classification, and graph 
convolutional networks \cite{c21}. Graph convolution 
can be implemented indirectly in the transform domain 
\cite{c19}. Define graph convolution as
$\mathbf{x}\star \mathbf{y}=\mathcal{F}^{-1}(\mathcal{F}(\mathbf{x})*\mathcal{F}(\mathbf{y}))$,
where $*$ denotes Hadamard product \cite{c12}, $\mathcal{F}$ is the graph Fourier transform (GFT) and $\mathcal{F}^{-1}$ 
denotes 
inverse GFT. 
This formulation exploits the property that graph
convolution in vertex
domain is equivalent to spectral domain product.

In \cite{c22}, a more general definition of graph convolution take 
the filtering form, i.e., 
$\mathbf{x}\star \mathbf{y}=f(\mathbf{x})=(\mathbf{V}diag(\mathbf{\hat y})\mathbf{V}^{-1})\cdot\mathbf{x}$,
where $\mathbf{V}$ consists of
eigenvectors of the representing matrix as columns
and $\hat{\mathbf{y}}$ can be interpreted as
filter parameters in graph spectral domain. 
Note that, these two definitions of graph convolution have the same mathematical formulation. For convenience, we define
spectral convolution over MLG
in a similar mathematical formulation as follows.
In M-GSP, the spectral convolution is defined as
\begin{equation}\label{conv3}
\mathbf{x}\star \mathbf{y}=\mathcal{F}_{\rm MLG}^{-1}(\mathcal{F}_{\rm MLG}(\mathbf{x})*\mathcal{F}_{\rm MLG}(\mathbf{y})),
\end{equation}
where $\mathcal{F}_{\rm MLG}$ is M-GFT (or M-GST) and 
$\mathcal{F}_{\rm MLG}^{-1}$ is the inverse M-GFT (or M-GST). Then, the MLG convolutional filter with parameters $\mathbf{c}$ on the signal $\mathbf{s}$ is defined by
$f_c(\mathbf{s})=\mathbf{s}\star\mathbf{c}$.

\subsection{Sampling and Interpolation}
\subsubsection{Sampling in Vertex Domain}
Within the M-GSP framework, sampling is
a process to select a subset of signals 
to describe some global information.
Consider the sampling of signal $\mathbf{s}\in\mathbb{R}^{M\times N}$ 
over MLG into $P$ layers with index 
$\{p_1,p_2,\cdots,p_P\}$ and $Q$ entities with index $\{q_1,q_2,\cdots,q_Q\}$. Let $\times_n$ be the $n$-mode product introduced in \cite{c12}.
The sampling operation is defined by
\begin{equation}\label{ds}
\mathbf{s}_{D}=\mathbf{s}\times_1\mathbf{S}_P\times_2\mathbf{S}_Q\in\mathbb{R}^{P\times Q}
\end{equation}
where sampling operator $\mathbf{S}_P\in\mathbb{R}^{P\times M}$ 
consists of elements
\begin{align}
[{S}_{P}]_{ij}=\left\{
\begin{array}{rcl}
1&\quad &j=p_i\\
0&\quad &\mbox{otherwise}
\end{array}, \right.
\end{align}
and $\mathbf{S}_Q\in\mathbb{R}^{Q\times N}$ with elements calculated as
\begin{align}
[{S}_{Q}]_{ij}=\left\{
\begin{array}{rcl}
1&\quad &j=q_i\\
0&\quad &\mbox{otherwise}
\end{array} .\right.
\end{align}
The interpolation operation is defined by 
\begin{equation}
\mathbf{s}_R=\mathbf{s}_D\times_1\mathbf{T}_M\times_2\mathbf{T}_N\in\mathbb{R}^{M\times N},
\end{equation}
where $\mathbf{T}_M\in\mathbb{R}^{M\times P}$ and $\mathbf{T}_N\in\mathbb{R}^{N\times Q}$.
Sampling in vertex domain can be viewed as a two-step sampling in layers and entities. M-GSP sampling and GSP sampling share many interesting properties. Readers may refer to \cite{c13} for more details.
\begin{figure*}[t]
	\centering
	\subfigure[Layer-Wise Sampling.]{
		\label{saml1}
		\includegraphics[height=2.5cm]{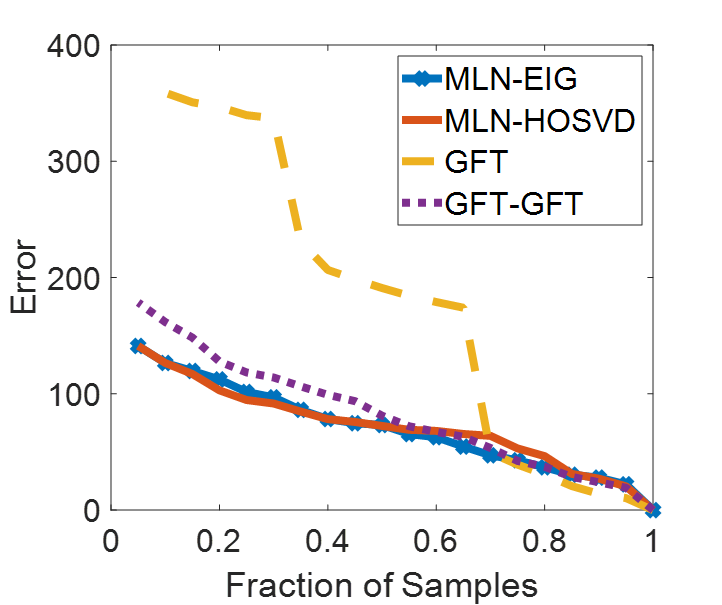}}
	\hfill
	\subfigure[Entity-Wise Sampling.]{
		\label{same1}
		\includegraphics[height=2.5cm]{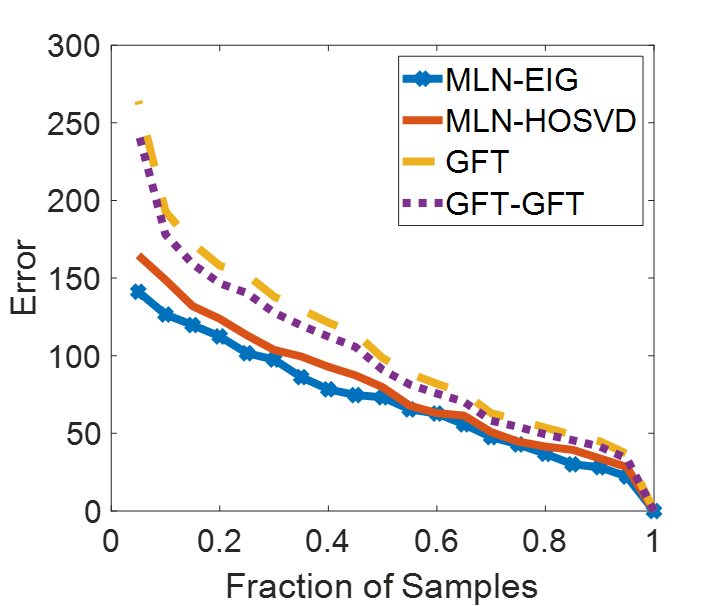}}
	\subfigure[Block-wise (Layer=3).]{
		\label{samb11}
		\hfill
		\includegraphics[height=2.5cm]{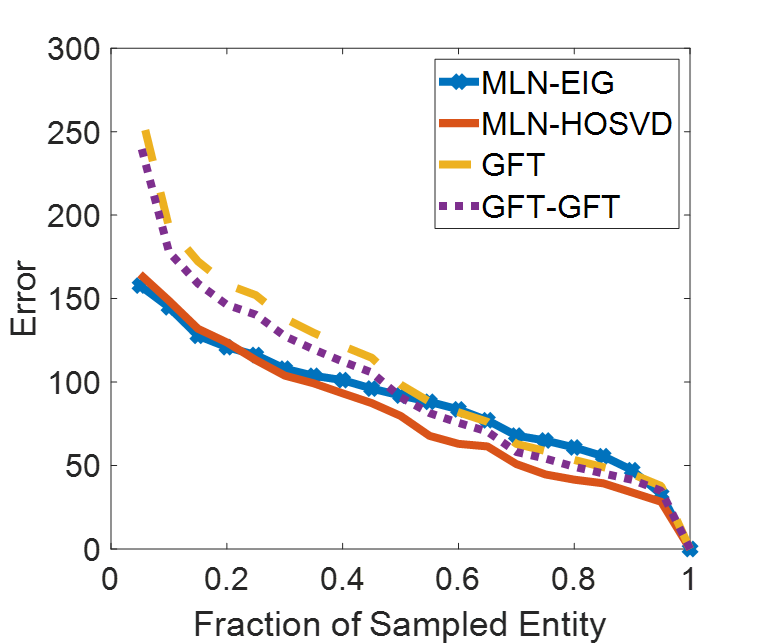}}
	\hfill
	\subfigure[Block-wise (Layer=2).]{
		\label{samb21}
		\includegraphics[height=2.5cm]{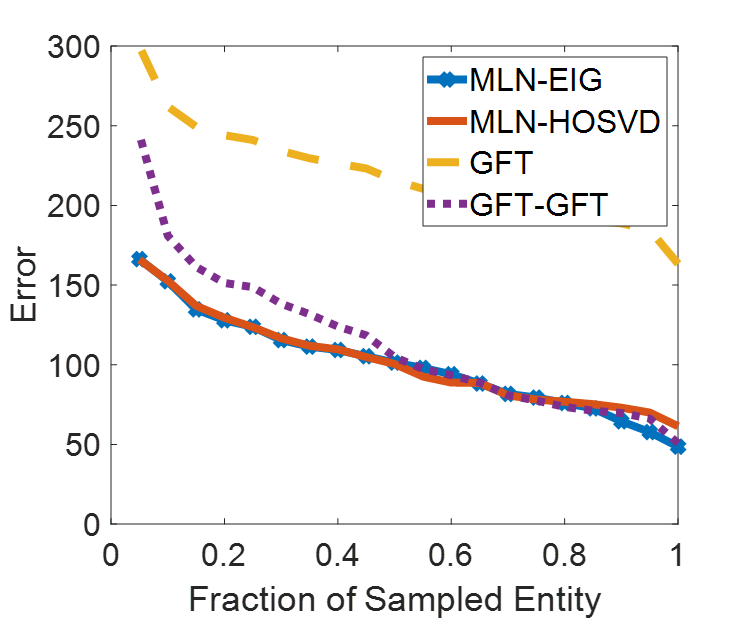}}
	\hfill
	\subfigure[Block-wise (Layer=1).]{
		\label{samb31}
		\includegraphics[height=2.5cm]{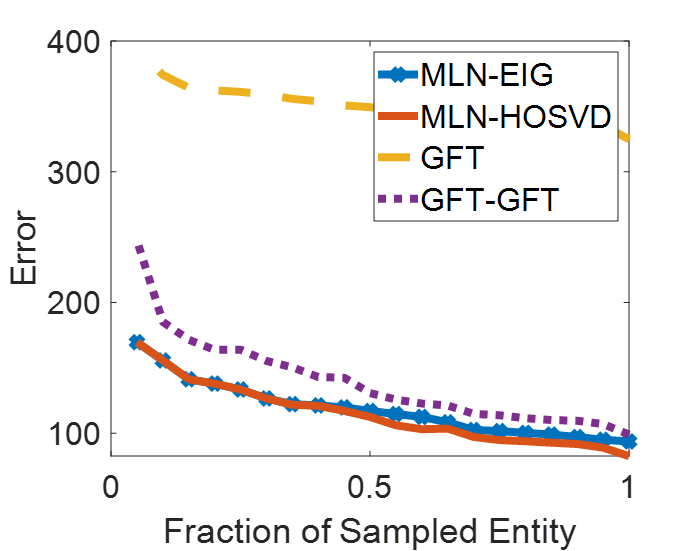}}
	\vspace{-1mm}
	\caption{Results of Sampling in the Order of Transformed Value.}
	\label{samex1}
	\vspace{-5mm}
\end{figure*}

\subsubsection{Sampling in Spectral Domain}\label{st}
Beyond direct sampling in the vertex domain, it can be more convenient and straightforward to sample signals in spectral domain. 
By neglecting spectral domain zeros, 
we can easily approximately recover 
the full signal from
samples by interpolating with zeros to 
the sampled signal transforms
before recovering with the inverse transform. Sampling in 
spectral domain is intuitive for lossless sampling and recovery.
Here we shall focus on the lossy sampling of tensor-based signals.

\begin{figure}[t]
	\centering
	\subfigure[Layer-Wise.]{
		\label{sam1}
		\includegraphics[height=1.5cm]{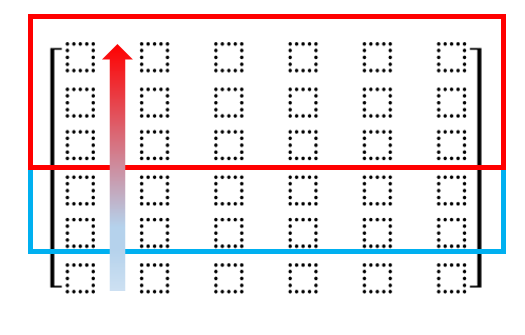}}
	\hfill
	\subfigure[Entity-Wise.]{
		\label{sam2}
		\includegraphics[height=1.5cm]{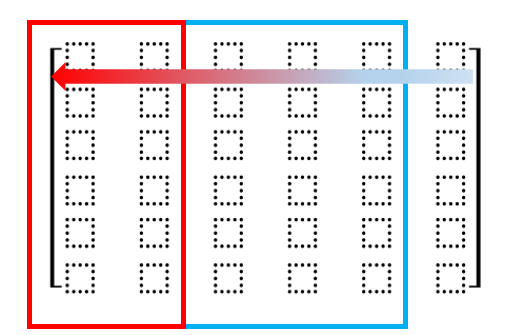}}
	\hfill
	\subfigure[Block-Wise.]{
		\label{sam3}
		\includegraphics[height=1.5cm]{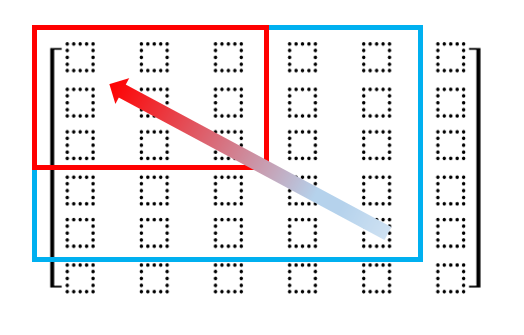}}
	\caption{Options of Sampling Directions for Tensor Signals.}
	\vspace{-3mm}
	\label{sam}
\end{figure}

Generally, we can sample signals in spectral domain according to Algorithm \ref{basic3}. 
We 
reshape transformed signals in terms of eigenvalues or transformed values from top left to bottom right in descending order. 
When sampling a sub-block in $\mathbf{\hat s}$, there are three different directions shown in Fig. \ref{sam}. Depending on the
structure and features of the underlying dataset, different methods 
may be suitable for different tasks. We will show examples of data compression based on the MLG sampling theory in Section \ref{app}.
\begin{algorithm}[htb]
	\begin{algorithmic}[1] 
		\caption{Sampling of Tensor Signal in Spectral Domain}\label{basic3}
		\STATE {\bf{Input}}: Signals $\mathbf{s}\in\mathbb{R}^{M \times N}$;
		\STATE Transform the signal into spectral space as $\mathbf{\hat s}\in\mathbb{R}^{M \times N}$ via M-GFT or M-GST;
		\STATE Reorder the elements in $\mathbf{\hat s}$ via certain rules;
		\STATE Sample transformed signal in the left top as $\mathbf{\hat s}_D\in\mathbb{R}^{P \times Q}$;
		\STATE Interpolate zeros in the right bottom of  $\mathbf{\hat s}_D$ as $\mathbf{\hat s}_R\in\mathbb{R}^{M \times N}$
		\STATE Implement the inverse MLG transform on $\mathbf{\hat s}_R$ to obtain the recovered signal $\mathbf{s}_R\in\mathbb{R}^{M \times N}$.
		\STATE  {\bf{Output}}: Sampled coefficients $\mathbf{\hat s}_D$ and recovered signal $\mathbf{s}_R$.
	\end{algorithmic}
\end{algorithm}

\section{Application Examples} \label{app}
\subsection{Image Compression}
Data compression through
sampling may reduce dataset size
for efficient storage and transmission. 
Moreover, projecting signals into a suitable manifold or
space, one may reveal a sparse representation of the data. In M-GSP, we can also select a subset of transformed signal samples in 
spectral domain to approximate the
full signal as suggested in Algorithm \ref{basic3}. 
Moreover, we can also integrate data encoding
to further compress data size. In this 
application example, we mainly focus on sampling 
to demonstrate the use of MLG spectrum space to represent 
structured data.

We test MLG sampling on the RGB icon dataset \cite{c14}. 
This dataset contains several icon images of size $\mathbb{R}^{16\times 16 \times 3}$. Our goal is to sample a subset of
$K=16\times 16 \times 3=768$ transformed data points
in spectral domain. We follow Algorithm \ref{basic3} to sample the transformed data before using the sampled data
to recover the original images. 
In this case, a 3-layer MLG is constructed where
the intralayer connections are formed by connecting to the 4-neighborhood pixels and the interlayer connections connect the counterparts of the same pixel in all layers.
We compare our methods with two GSP-based methods. 
The first method (GFT) builds a graph with $16\times 16$ nodes and implements GFT on signals in each layer separately. The second method (GFT$^2$) builds a graph of size $16\times 16$ for 
pixels that are neighbor-4 connected
and constructs a graph of size $3$ for (fully connected) frames. 
Let $\mathbf{E}$ be the spectrum of frame-wise graph, and $\mathbf{F}$ be the spectrum of pixel-wise graph.
We apply GFT on the 2-D signal 
$\mathbf{s}\in\mathbb{R}^{3 \times 256}$ 
to generate
$\mathbf{\hat s}=\mathbf{E}^{\mathrm{T}}\mathbf{s}\mathbf{F}.$
For both GSP and M-GSP methods, we apply the Laplacian matrix/tensor. 
We order the transformed signal in descending order of 
the transformed values from top left to bottom right, i.e., 
keeping larger values in Fig. \ref{samex1}.
As mentioned in Section \ref{st}, there are three directions for sampling node values. {In layer-wise sampling, 
	we remove samples layer-by-layer from bottom to
	up. More specifically, we remove samples from right to left when sampling each layer. Similarly, in  entity-wise sampling, we remove samples from right to left. When sampling each column, we move {from bottom to top}. In
	block-wise sampling, we preserve
	samples in top-left blocks. Since there are only three layers, we present all results of block-wise sampling layer-by-layer in Fig. \ref{samb11}-\ref{samb31}.}
	
Defining {\em sampling fraction}
as the ratio of saved samples to 
total original signal samples, 
we measure errors
between the recovered and original images. 
The results from 
are shown in Fig. \ref{samex1}, where MLN-EIG and MLN-HOSVD are from MLG Fourier and singular space respectively. 
These results confirm that the 
proposed MLG methods 
provide better performance than GSP.

\subsection{Edge Detection}
\begin{figure}[t]
	\centering
	\subfigure[2-D.]{
		\label{dk1}
		\includegraphics[height=1.2cm]{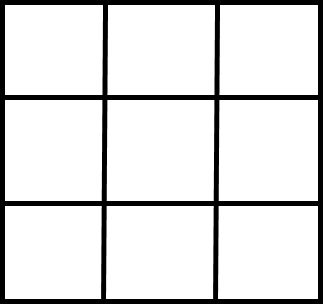}}
	\hspace{2cm}
	\subfigure[3-D.]{
		\label{dk2}
		\includegraphics[height=1.2cm]{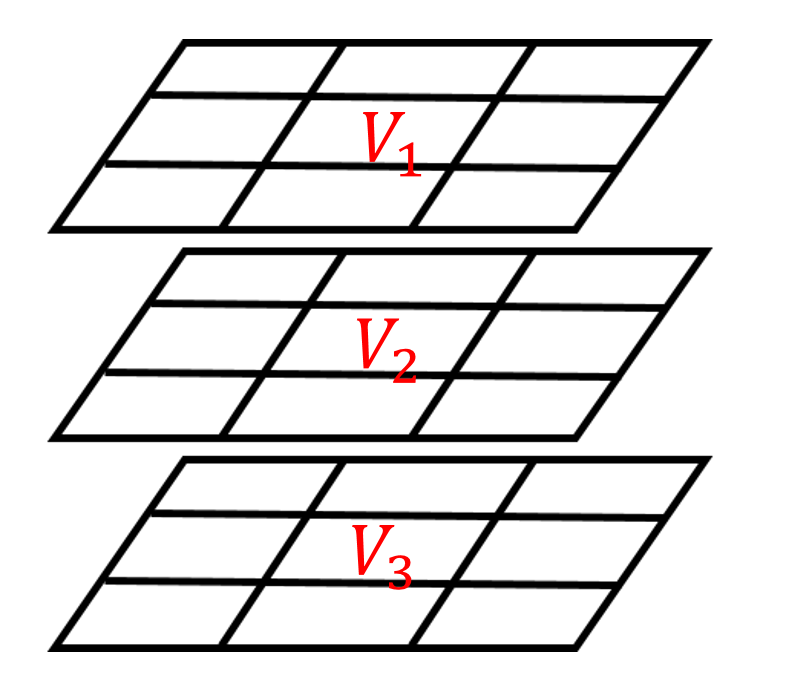}}
	\vspace{-2mm}
	\caption{Example of Convolution Kernels.}
	\vspace{-5mm}
	\label{dk}
\end{figure}

Convolutions are
widely used for edge detection of images. In traditional processing, one may first transform RGB to gray-scale
images for edge detection using operators such as Prewitt, Canny, and Sobel \cite{c42,c43}. Similarly, geometric convolution methods could also detect edges in gray-scale images \cite{c20}. 
Unlike
traditional (graph) signal processing,
M-GSP can model RGB images by using 3-D windows/blocks and define 3-D 
convolution kernels as shown in Fig. \ref{dk2}.
{To construct an 
MLG for edge detection, we also 
represent each 
RGB color by a layer and
connect each pixel in one layer with its counterparts in other layers for interlayer connections.}
Each pixel is connected with its four adjacency neighbors for intralayer connections.
The benefit of constructing such graph is to 
bypass the computation needed
to regenerate spectrum for each 
window (block). For a $3\times 3 \times 3$ kernel, we construct a 3-layer MLG with 9 nodes in each layer.

\begin{figure}[t]
	\centering
	\includegraphics[width=3.5in]{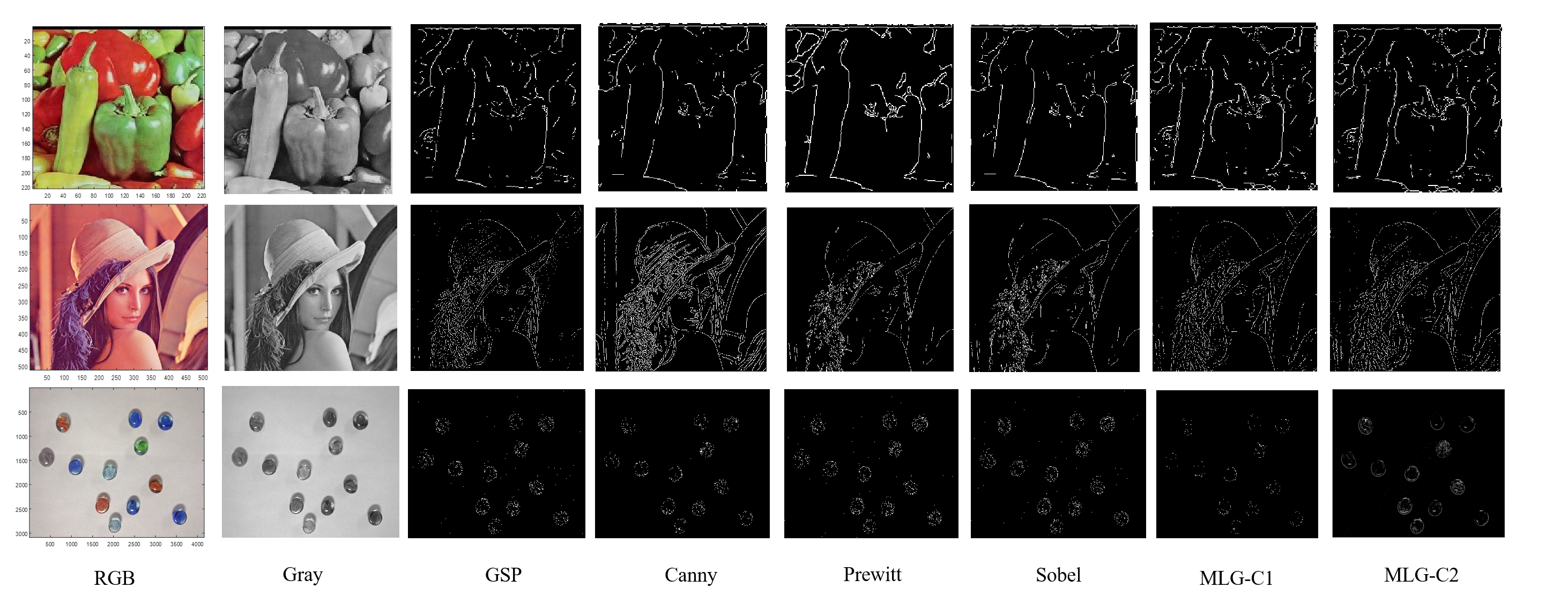}
	\vspace{-3mm}
	\caption{Results of Edge Detection.}
	\label{flat}
	\vspace{-5mm}
\end{figure}

In general, one can first extract smooth image features, before
applying edge detection on the
difference between original and 
smoothed images. To define a graph localization kernel for smoothing, we start from the single-layer graph.
Let $K$ be the number of nodes and
the graph convolution kernel be
$\mathbf{c}\in\mathbb{R}^K$ where
$c_{i}=1$ for $i=k_1,\cdots,k_z$ with $z\leq K$; otherwise, $c_i=0$.
The convolution output for a signal $\mathbf{s}\in\mathbb{R}^K$ is
$\mathbf{s'}=\mathbf{V}[(\mathbf{V}^\mathrm{T}\mathbf{s})*(\mathbf{V}^\mathrm{T}\mathbf{c})]
=\sum_j(\mathbf{f}_j\mathbf{f}_j^\mathrm{T}\mathbf{s}\sum_i[f_j]_{k_i})$,
where $\mathbf{V}=[\mathbf{f}_1,\cdots,\mathbf{f}_K]$ is the graph spectrum. 
Different from a pure shift,
the convolution kernel enhances
information around nodes $\{v_{k_i}\}$.
With locally-enhanced gray-scale images, we can replace centroid of a window by the mean of convolution outputs inside that window
to smooth signals further. By shifting the window over the entire image, one can obtain a locally-enhanced blurred
gray-scale image.

Similarly, since our goal is 
to smooth signals 
within the 3-D window in MLG, we can simply localize the information around the center of the block. Here, we design two 3-D kernels with size $3\times k\times k$, i.e, $\mathbf{c}_1\in\mathbb{R}^{3\times k^2}$ with entry in node $v_2$ as 1 and all
other entries as zero, and $\mathbf{c}_2\in\mathbb{R}^{3\times k^2}$ with entries in node $v_1,v_2,v_3$ as 1 and all other entries as zero, indexed as \ref{dk2}. We then obtain a smoothed
gray-scale image from convolution output and determine
its difference from the original gray-scale image generated from RGB. A threshold is designed here to detect edges based on the explicit difference between the smoothed image and its original. MLG spectra are applied here to implement the MLG convolution.
Results from the MLG edge detection can be visually examined
in Fig. \ref{flat}, compared with 
results from several classic detection methods (e.g., Canny, Prewitt, Sobel) and a GSP-based method.  The comparison shows that
MLG-based methods 
yield more smooth edges and show
clearer details.
\begin{figure}[htb]
	\centering
	\includegraphics[width=3.2in]{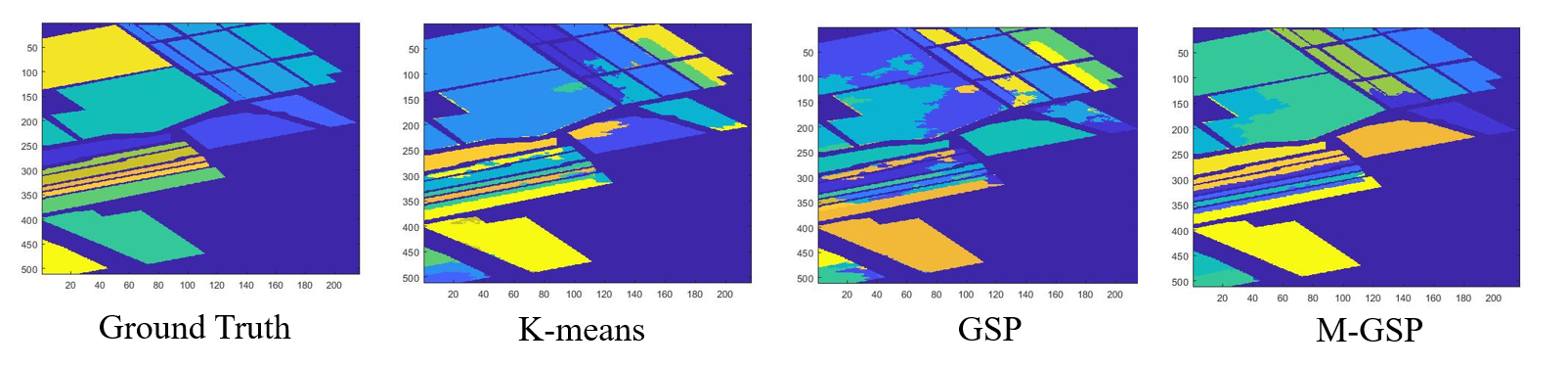}
	\vspace{-3mm}
	\caption{Results of HSI Segmentation.}
	\label{flat11}
	\vspace{-5mm}
\end{figure}

\vspace{-3mm}
\subsection{Hyperspectral Image Segmentation}
Spectral clustering \cite{d1} is a widely-used method for unsupervised image segmentation. In this part, we introduce unsupervised hyperspectral image (HSI) segmentation based on MLG singular space. To construct a MLG, we first cluster the multi-spectral frames into $M=10$ clusters as layers and construct $N=100$ superpixels as entities. The new signals in each layer is the mean of all frames in that cluster. By constructing intralayer and interlayer connections using Gaussian distance, we generate an $M$-layer MLG with $N$ nodes in each layer. 
Implementing HOSVD on the adjacency tensor, we can obtain the entity-wise (superpixel) singular vectors in descending order of the entity-wise singular values, i.e., $\mathbf{E}_e=[\mathbf{e}_1\cdots\mathbf{e}_N]\in\mathbb{R}^{N\times N}$. Selecting the first $P$ singular tensors as $\mathbf{P}_k=[\mathbf{e}_1,\cdots,\mathbf{e}_P]\in\mathbb{R}^{N\times P}$ based on the largest singular gap, we can cluster the rows of $\mathbf{P}_k$ into $Q$ groups based on $k$-means clustering. The superpixel $i$ is clustered into group $j$ if the $i$-th row of $\mathbf{P}_k$ is in group $j$. Labeling the pixels as the same cluster of its superpixel, we segment the HSI. Let us compare the M-GSP spectral clustering with $k$-means and GSP spectra clustering in datasets named Indian Pine, Pavia University and Salinas. The visualization results of the segmented Salinas HSI are shown as Fig. \ref{flat11}, where M-GSP illustrates more robust results. In addition to visual results, we also
consider quantitative metrics. We compute the edges of segmentation results from different methods, and compare them to the ground truth in Table \ref{accs}. M-GSP continues to show superior performance in all HSIs.
\begin{table}[htb]	\vspace{-2mm}

	\caption{Accuracy of Segmentation Boundaries}
	\centering
	\begin{tabular}{|l|l|l|l|}
		\hline
		Data    & $k$-means & GSP    & M-GSP  \\ \hline
		IndianP & 0.8257  & 0.8298 & \textbf{0.8441} \\ \hline
		Salinas & 0.9208  & 0.9285 & \textbf{0.9409} \\ \hline
		PaviaU  & 0.9070  & 0.9088 & \textbf{0.9255} \\ \hline
	\end{tabular}
	\label{accs}
	\vspace{-4mm}
\end{table}
\section{Conclusions}
In this work, we introduce image modeling based on multilayer graphs. We present the fundamentals of M-GSP frequency analysis together with the spectral operations. We further provide several example applications based on M-GSP operations. Our test results demonstrate the efficacy and strong 
future potentials of applying M-GSP in image processing. Other applications of M-GSP 
may include
dynamic point clouds and video signals.


\begin{thebibliography}{34}
\bibitem{c1} G. Cheung, E. Magli, Y. Tanaka and M. K. Ng, ``Graph spectral image processing," in \textit{Proceedings of the IEEE}, vol. 106, no. 5, pp. 907-930, May 2018.

\bibitem{c2} A. Ortega, P. Frossard, J. Kovačević, J. M. F. Moura and P. Vandergheynst, "Graph signal processing: overview, challenges, and applications," in \textit{Proceedings of the IEEE}, vol. 106, no. 5, pp. 808-828, May 2018.

\bibitem{c3} G. Fracastoro, D. Thanou and P. Frossard, ``Graph transform optimization with application to image compression," in \textit{IEEE Transactions on Image Processing}, vol. 29, pp. 419-432, Aug. 2020.

\bibitem{c4} M. Onuki, S. Ono, M. Yamagishi and Y. Tanaka, ``Graph signal denoising via trilateral filter on graph spectral domain," in \textit{IEEE Transactions on Signal and Information Processing over Networks}, vol. 2, no. 2, pp. 137-148, Jun. 2016.

\bibitem{c5} M. D. Domenico, A. Solé-Ribalta, E. Cozzo, M. Kivelä, Y. Moreno, M. A. Porter, S. Gómez, and A. Arenas, ``Mathematical formulation of multilayer networks," \textit{Physical Review X}, vol. 3, no. 4, p. 041022, Dec. 2013.

\bibitem{c6} P. Das and A. Ortega, ``Graph-based skeleton data compression," \textit{2020 IEEE 22nd International Workshop on Multimedia Signal Processing (MMSP)}, Tampere, Finland, Sep. 2020, pp. 1-6.

\bibitem{c7} F. Grassi, A. Loukas, N. Perraudin and B. Ricaud, ``A time-vertex signal processing framework: scalable processing and meaningful representations for time-series on graphs," in \textit{IEEE Transactions on Signal Processing}, vol. 66, no. 3, pp. 817-829, Feb. 2018

\bibitem{c8} J. S. Stanley, E. C. Chi and G. Mishne, ``Multiway graph signal processing on tensors: integrative analysis of irregular geometries," in \textit{IEEE Signal Processing Magazine}, vol. 37, no. 6, pp. 160-173, Oct. 2020.

\bibitem{c9} S. Zhang, Q. Deng, and Z. Ding. ``Introducing Graph Signal Processing over Multilayer Networks: Theoretical Foundations and Frequency Analysis", arXiv:2108.13638, 2022. [available online: \url{https://github.com/zsy93/Introducing-Graph-Signal-Processing-over-Multilayer-Networks/blob/main/ms.pdf}]

\bibitem{c10} A. Afshar, J. C. Ho, B. Dilkina, I. Perros, E. B. Khalil, L. Xiong, and V.
Sunderam, ``Cp-ortho: an orthogonal tensor factorization framework for
spatio-temporal data," in \textit{Proceedings of the 25th ACM SIGSPATIAL International Conference on Advances in Geographic Information Systems},
Redondo Beach, CA, USA, Jan. 2017, p. 67.

\bibitem{c11} L. De Lathauwer, B. De Moor, and J. Vandewalle, ``A multilinear singular value decomposition," \textit{SIAM Journal on Matrix Analysis and Applications}, vol. 21, no. 4, pp. 1253-1278, Jan. 2000.

\bibitem{c21} T. N. Kipf and M. Welling,
``Semi-supervised classification with graph convolutional networks," \textit{ICLR}, Toulon, France, Apr. 2017.

\bibitem{c19} D. I. Shuman, S. K. Narang, P. Frossard, A. Ortega and P. Vandergheynst, ``The emerging field of signal processing on graphs: Extending high-dimensional data analysis to networks and other irregular domains," in \textit{IEEE Signal Processing Magazine}, vol. 30, no. 3, pp. 83-98, May 2013.

\bibitem{c22} J. Shi, and J. M. Moura, ``Graph signal processing: Modulation, convolution, and sampling," \textit{arXiv preprint arXiv:1912.06762}, Dec. 2019.

\bibitem{c12} T. G. Kolda and B. W. Bader, ``Tensor decompositions and applications,"
\textit{SIAM Review}, vol. 51, no. 3, pp. 455-500, Aug. 2009.

\bibitem{c13} S. Chen, R. Varma, A. Sandryhaila and J. Kovačević, ``Discrete signal processing on graphs: sampling theory," in \textit{IEEE Transactions on Signal Processing}, vol. 63, no. 24, pp. 6510-6523, Dec. 2015.

\bibitem{c14} S. Zhang, Z. Ding and S. Cui, ``Introducing hypergraph signal processing: theoretical foundation and practical applications," in \textit{IEEE Internet of Things Journal}, vol. 7, no. 1, pp. 639-660, Jan. 2020.

\bibitem{c42} A. K. Cherri, and M. A. Karim, ``Optical symbolic
substitution: edge detection using Prewitt, Sobel, and
Roberts operators," \textit{Applied Optics}, vol. 28, no. 21, pp.
4644-4648, Nov. 1989.

\bibitem{c43} L. Ding, and A. Goshtasby, ``On the Canny edge detector," \textit{Pattern Recognition}, vol. 34, no. 3, pp. 721-725, Mar. 2001.

\bibitem{c20} S. Zhang, S. Cui and Z. Ding, ``Hypergraph-based image processing,"
\textit{2020 IEEE International Conference on Image Processing (ICIP)}, Abu
Dhabi, United Arab Emirates, Oct. 2020, pp. 216-220.

\bibitem{d1} U. Von Luxburg, ``A tutorial on spectral clustering," \textit{Statistics and computing}, vol. 17, no. 4, pp. 395-416, Aug. 2007.
\end{thebibliography}
\end{document}